\documentclass[aps,prd,preprint,superscriptaddress]{revtex4}
\usepackage{graphicx}
\usepackage{amsmath}
\usepackage{bm}% bold math
\usepackage{color}% bold math

\def\ee{\end{eqnarray}}

\def\=:{=\hspace{-.7em}\raisebox{1.1ex}{.}\hspace{.1em}\raisebox{-0.2ex}{.} }
\def\ee{\end{eqnarray}}

\def\=:{=\hspace{-.7em}\raisebox{1.1ex}{.}\hspace{.1em}\raisebox{-0.2ex}{.} }

\newcommand {\beq}{\begin{eqnarray}}
\newcommand {\eeq}{\end{eqnarray}}
\newcommand {\non}{\nonumber\\}

\newcommand {\1}[1]{\frac{1}{#1}}

\newcommand {\del}{\partial}

\begin{document}

%\preprint{}

\title{Jewels on a wall ring\\
(Sine-Gordon kinks on a domain wall ring)
}


\author{Michikazu Kobayashi}
\affiliation{
Department of Basic Science, University of Tokyo, Komaba 3-8-1, Meguro-ku, Tokyo 153-8902, Japan
}
\author{Muneto Nitta}
\affiliation{Department of Physics, and Research and Education Center for Natural 
Sciences, Keio University, Hiyoshi 4-1-1, Yokohama, Kanagawa 223-8521, Japan
}
%\thanks{}



\date{\today}
\begin{abstract}
We construct a stable domain wall ring with lump beads on it 
in a baby Skyrme model with a potential consisting of two terms linear and quadratic in fields.

\end{abstract}
\pacs{}
%\keywords{}

\maketitle

\section{Introduction}

Vortices and domain walls are topological solitons present in various physical systems from field theory \cite{Manton:2004tk}
 and cosmological models \cite{Vilenkin:2000} 
to condensed matter systems \cite{Volovik:2003}.  
In contrast to instantons and monopoles which have been studied 
extensively in high energy physics and mathematics,   
vortices and domain walls have not been paid much attention thus far 
in those fields.
However they play essential roles in condensed matter systems 
such as superconductors, superfluids, magnetism, quantum Hall states, 
nematic liquids, optics, and so on.  
The coexistence of these two kinds of solitons can also happen in 
various condensed matter systems;
a Bloch line in a Bloch wall in magnetism \cite{Chen:1977}, 
half-quantized vortices inside a chiral domain wall 
in chiral $p$-wave superconductors \cite{Garaud:2012}, 
and a Mermin-Ho vortex within a domain wall in 
$^3$He superfluid (see Fig.~16.9 of Ref.~\cite{Volovik:2003}). 
Further examples can be found in the limit of infinitely heavy domain walls:  
Josephson vortices within an insulator in Josephson junctions 
of two superconductors \cite{Ustinov:1998} 
and
Josephson vortices in high-$T_c$ superconductors 
with multi-layered structures \cite{Blatter:1994} and in 
two coupled Bose-Einstein condensates \cite{Kaurov:2005}, 
where the insulators or inter-layers can be regarded as (heavy) domain walls. 
In all these cases, vortices become sine-Gordon solitons 
once absorbed into a domain wall.
A field theoretical model of 
the coexistence of domain walls and vortices as Josephson vortices 
was given recently \cite{Nitta:2012xq} 
in order to explain a previously known relation between vortices and sine-Gordon solitons \cite{Sutcliffe:1992ep}.
(They are the lowest dimensional example of 
``matryoshka Skyrmions" \cite{Nitta:2012wi}.)
Slightly different field theoretical models admitting the coexistence of domain walls and vortices were also considered before \cite{Ritz:2004mp}.

In these cases, vortices are all absorbed into a domain wall. 
In this sense, there seem to be no freely moving vortices in the bulk 
outside the domain wall.
However, a question arises. What happens if one makes a closed loop of the domain wall? If a vortex is absorbed into a closed domain line, it may be regarded as a freely moving vortex, apart from its stability.
In fact, recently such configurations of a domain wall ring with vortices on it 
have been theoretically proposed in condensed matter systems, 
such as chiral $p$-wave superconductors \cite{Garaud:2012} and 
multi-gap superconductors \cite{Garaud:2012pn}.
Motivated by these works, in this paper, we propose a field theoretical model 
admitting a stable domain wall ring with vortices absorbed in it. 

We consider an $O(3)$ nonlinear sigma model 
on the target space $S^2$ in $d=2+1$ dimensions, 
described by a unit three-vector of scalar fields ${\bf n}(x)=(n_1(x),n_2(x),n_3(x))$  
with the constraint ${\bf n}^2=1$, 
which is equivalent to a ${\bf C}P^1$ model.
The $O(3)$ model admits 
lumps or sigma model instantons \cite{Polyakov:1975yp} 
as a relative of vortices. 
The ${\bf C}P^1$ model with a potential term admitting two discrete vacua is known as the massive  ${\bf C}P^1$ model,
which can be made 
supersymmetric with additional fermions 
\cite{AlvarezGaume:1983ab} 
and admits a Bogomol'nyi-Prasad-Sommerfield 
domain wall solution 
interpolating the two discrete vacua  \cite{Abraham:1992vb,Arai:2002xa}.
In the presence of a potential term,  
lumps are unstable to shrink in general. 
Instead, if one gives them a linear time-dependence on 
their $U(1)$ moduli, they become stable Q-lumps \cite{Leese:1991hr}.  
If we consider a four derivative (Skyrme) term, the lumps are stabilized 
to become baby Skyrmions \cite{Piette:1994ug}. 

A closed domain line or a wall ring 
is nothing but a lump 
if the $U(1)$ modulus of the domain wall is twisted 
along the ring \cite{Nitta:2012kj}. 
This twisted domain wall ring is unstable to shrink unless 
one puts linear time-dependence on the $U(1)$ modulus 
or adds the Skyrme  term, as denoted above.   
More precisely, the originally proposed 
baby Skyrme model has the potential term 
$V = m^2 (1-n_3)$ \cite{Piette:1994ug},  
which admits the unique vacuum and does not admit 
a domain wall.
A new baby Skyrme model proposed later \cite{Weidig:1998ii,Kudryavtsev:1997nw} has the potential
$V = m^2 (1-n_3)(1+n_3)$ 
admitting two discrete vacua $n_3=\pm 1$ 
and a domain wall 
solution interpolating between these two vacua 
\cite{Kudryavtsev:1997nw,Harland:2007pb}, 
as the case without a Skyrme term \cite{Abraham:1992vb,Arai:2002xa}.
In this model, a baby Skyrmion is in fact in 
the shape of a domain wall ring.
In this paper, we consider both types of the potential terms 
$V = \beta^2 n_1 + m^2 (1-n_3)(1+n_3)$ 
in the regime $\beta \ll m$.  
In magnetism, this potential term appears in 
Heisenberg ferromagnets with two easy axes.
Such an $O(3)$ sigma model without the Skyrme term
was studied recently to consider a vortex absorbed into 
a straight domain wall \cite{Nitta:2012xq}, 
but it does not admit a stable domain wall ring. 
Here we consider the Skyrme term to stabilize a domain wall ring.
We numerically construct domain wall rings with 
one, two and three vortices (lumps), 
which have the topological lump charges $k=1,2,3$, respectively, 
looking like jewels on a ring. 
These vortices are sine-Gordon kinks on the domain wall ring. 
We find that lumps are placed with the same distance from each other 
because of repulsions among them. 

This paper is organized as follows. 
After our model is given in Sec.~\ref{sec:model},
we give a numerical solution of a twisted domain line 
in the absence of the term $\beta^2 n_1$ in the potential
in Sec.~\ref{sec:closed-wall}. 
It carries the topological lump charge 
and is nothing but a baby Skyrmion.
In Sec.~\ref{sec:wall},  we give  
numerical solutions of a lump within a straight domain wall 
in the models with the term $\beta^2 n_1$ in the potential 
without \cite{Nitta:2012xq}  and with the Skyrme term.
Then, in Sec.~\ref{sec:wall-ring}, 
we give numerical solutions of domain wall rings with 
one, two and three lumps. 
Section \ref{sec:summary} is devoted to a summary 
and discussion.

%%%%%%%%%%%%%%%%%%%%%%%%%%
\section{The model \label{sec:model}}
We consider an $O(3)$ sigma model in $d=2+1$ dimensions
described by a three vector of scalar fields 
${\bf n} (x)= (n_1(x),n_2(x),n_3(x))$  
with a constraint ${\bf n} \cdot {\bf n} = 1$.
The Lagrangian of our model is given by
\beq
&& {\cal L} = \1{2} \del_{\mu}{\bf n}\cdot \del^{\mu} {\bf n} 
 - {\cal L}_4({\bf n})
 - V({\bf n}), \quad  
 \label{eq:Lagrangian}
\eeq
with $\mu=0,1,2$. 
Here, the four derivative (baby Skyrme) term is expressed as
\beq
{\cal L}_4 ({\bf n})
= \kappa  \left[{\bf n} \cdot 
 (\partial_{\mu} {\bf n} \times \partial_{\nu} {\bf n} )\right]^2
= \kappa (\partial_{\mu} {\bf n} \times \partial_{\nu} {\bf n} )^2 , 
\eeq
and the potential term is given by
\beq
V({\bf n}) = m^2(1-n_3^2) + \beta^2 n_1.\label{eq:pot}
\eeq 
The potential with $m=0$ was originally considered in Ref.~\cite{Piette:1994ug}, 
and the one with $\beta=0$ was proposed later in 
Refs.~\cite{Weidig:1998ii,Kudryavtsev:1997nw}.
The choice of our potential is physically quite natural, since 
it is known as the Heisenberg ferromagnet with 
anisotropy with two easy axes.

With introducing the projective coordinate $u (\in {\bf C})$ of ${\bf C}P^1$ 
by
\beq
 n_i = \phi^\dagger \sigma_i \phi, \quad
 \phi^T = (1, u)/\sqrt{1+|u|^2},
\eeq
the Lagrangian (\ref{eq:Lagrangian}) 
can be rewritten in the form of the ${\bf C}P^1$ model 
with potential terms, given by
\beq
&& {\cal L} = 
2\frac{\partial_{\mu} u^* \partial^{\mu} u}
  {(1 + |u|^2)^2} 
- 8 \kappa \frac{(\del_{\mu}u^* \del^{\mu}u)^2 
 - |\del_{\mu} u \del^{\mu} u|^2}
 {(1+|u|^2)^4} 
 - V \label{eq:CP1} \\ 
&& V= m^2 (1-D_3^2) +\beta^2 D_1 
= m^2 g^{uu^*} |\partial_u D_3|^2 +\beta^2 D_1 , \quad \\
&& 
 D_3 \equiv  \frac{1 - |u|^2}{1 + |u|^2} = n_3, \quad
 D_1 \equiv \frac{u + u^\ast}{1+|u|^2} = n_1 . 
\eeq
Here, $g_{uu^*}=1/(1+|u|^2)^2$ is the K\"ahler (Fubini-Study) metric 
of  ${\bf C}P^1$,  
$g^{uu^*}=(1+|u|^2)^2$ is its inverse, 
and 
$D_i = n_i$ are called the moment maps (or the Killing potentials) 
of the $SU(2)$ isometry 
generated by $\sigma_i$, respectively. 
With $\beta =0$ and $\kappa=0$, this model is known as the massive 
${\bf C}P^1$ model 
with the potential term of the norm of the Killing vector $\del_u D_3$ 
corresponding to the isometry generated by $\sigma_3$,
which is a truncated version of 
a supersymmetric sigma model with eight supercharges \cite{AlvarezGaume:1983ab,Abraham:1992vb}.

%%%%%%%%%%%%%%%%%%%%%%%%%%%%%%%%%%%%%%%%%%%%%%%%%%
\section{Twisted closed domain wall \label{sec:closed-wall}}

For a while, let us ignore the four derivative baby Skyrme term 
($\kappa=0$) in the Lagrangian in Eq.~(\ref{eq:Lagrangian}).
Let us consider the potential in Eq.~(\ref{eq:pot}) with $\beta=0$.
It admits two discrete vacua $n_3 = \pm 1$. 
A domain wall or an anti-domain wall solution interpolating these two vacua 
is given by \cite{Abraham:1992vb,Arai:2002xa,Nitta:2012wi}
\beq
&& \theta (x^1) = 2 \arctan \exp (\pm \sqrt 2 m (x^1 -X)),  
  \quad 0 \leq \theta \leq \pi , \nonumber \\
&&  n_1 = \cos \alpha \sin \theta (x^1) , \quad 
      n_2 = \sin \alpha  \sin \theta (x^1) , \quad
      n_3 = \cos \theta (x^1),
\label{eq:wall} 
\eeq
with a phase modulus 
$\alpha$ ($0 \leq \alpha < 2\pi$) and 
the translational modulus $X \in {\bf R}$ 
of the domain wall. 
The moduli $\alpha$ and $X$ can be regarded as 
Nambu-Goldstone modes corresponding to 
$U(1)$ and translational symmetries spontaneously broken down 
in the vicinity of the domain wall, respectively. 
A domain wall solution in the presence of the baby Skyrme term was studied in Refs.~\cite{Kudryavtsev:1997nw,Harland:2007pb}.

\begin{figure}[h]
\begin{center}
\includegraphics[width=0.9\linewidth,keepaspectratio]{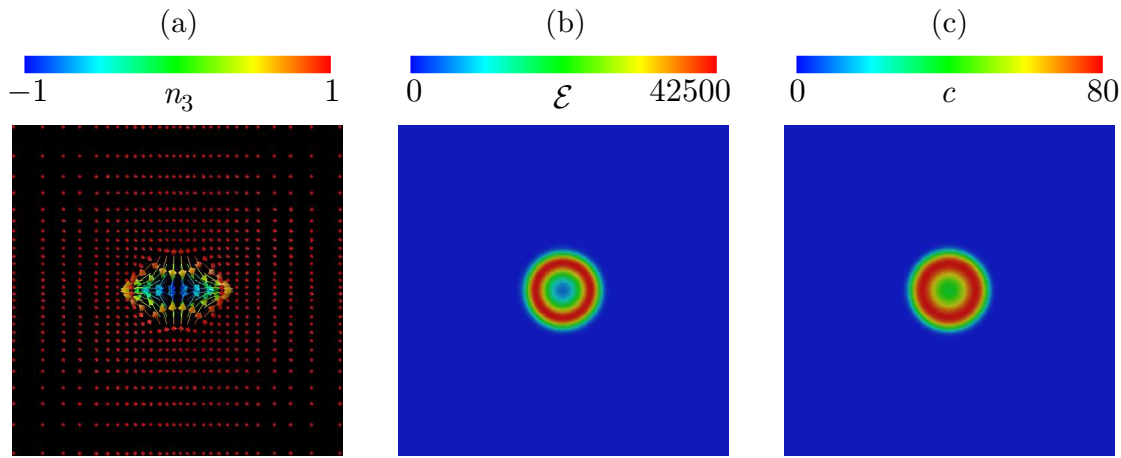}
\caption{A twisted domain wall ring as a baby Skyrmion.
(a) The textures ${\bf n}(x)$. The color of each arrow shows the value of $n_3$.
(b) The total energy density $\mathcal{E} \equiv ( \del_a{\bf n}\cdot \del^a {\bf n} ) / 2 + {\cal L}_4({\bf n}) + V({\bf n})$ ($a=1,2$).
(c) The topological lump charge density $c \equiv  \{ {\bf n} \cdot (\partial_1 {\bf n} \times \partial_2 {\bf n}) \} / (4 \pi)$.
As numerical parameters, we fix $\kappa = 0.02$ and $m^2 = 20000$, and plot the values in the region $-0.29 \leq x^{1,2} \leq 0.29$.
\label{fig:wall-ring0}}
\end{center}
\end{figure}

A loop of the domain wall carries a lump charge 
if the $U(1)$ modulus $\alpha$ winds along the wall loop 
\cite{Nitta:2012kj}. 
The topological charge of the lump $\pi_2(S^2) \simeq {\bf Z}$ 
is given by
\beq
 k 
&=& \1{4 \pi}  \int d^2x\: {\bf n}\cdot 
(\partial_1 {\bf n} \times \partial_2 {\bf n} )
=
 \1{4 \pi}  \int d^2x\:
\epsilon_{ijk}
n_i \partial_{1} n_j  \partial_{2} n_k  \non \label{eq:lump-charge}
&=& \1{2 \pi}  \int d^2x 
{i(\del_1 u^* \del_2 u - \del_2 u^* \del_1 u)
\over (1+|u|^2)^2} .
\eeq
However, a twisted closed wall line is unstable to shrink. 
It can be stabilized in the presence of the baby Skyrme term, 
which results in a baby Skyrmion.  
%%%%%%%%%%%%%%%%
%%%%%%%%%%%%%%%%
We construct a numerical solution of a twisted 
domain wall ring with the unit lump charge 
($k=1$)
by a relaxation method;
see Fig.~\ref{fig:wall-ring0}. 
%%%%%%%%%%%%%%%%%%%%%%
%%%%%%%%%%%%%%%%%%%%%%
One can see that the topological lump charge as well as the energy density is uniformly 
distributed along the ring.

In the context of magnetism, this configuration is called a bubble domain 
\cite{Chen:1977}.

%%%%%%%%%%%%%%%%%%%%%%%%%%%%%%%%%%%%%%%%%%%%%%%%%%
\section{Sine-Gordon kinks on a straight domain wall 
\label{sec:wall}}
With promoting the moduli to fields $\alpha(t,x^2)$ and $X(t,x^2)$ 
on the domain wall world-volume $(t,x^2)$, 
the effective theory of the domain wall can be constructed 
by the moduli approximation \cite{Manton:1981mp}. 
It is a free field theory of $\alpha(t,x^2)$ and $X(t,x^2)$ 
or a sigma model on ${\bf R} \times S^1$.
The $U(1)$ symmetry is explicitly broken when $\beta$ is 
taken into account in the potential (\ref{eq:pot}) \cite{footnote}. 
Then, a potential term is induced on the domain wall effective action 
and it becomes the sine-Gordon model 
\cite{Nitta:2012xq}. 
A sine-Gordon kink in the wall effective theory 
corresponds to a lump in the bulk \cite{Nitta:2012xq}, 
in which the topological lump charge $k$ coincides with 
the topological charge $k$ of sine-Gordon kinks. 

\begin{figure}%[h]
\centering
\includegraphics[width=0.6\linewidth]{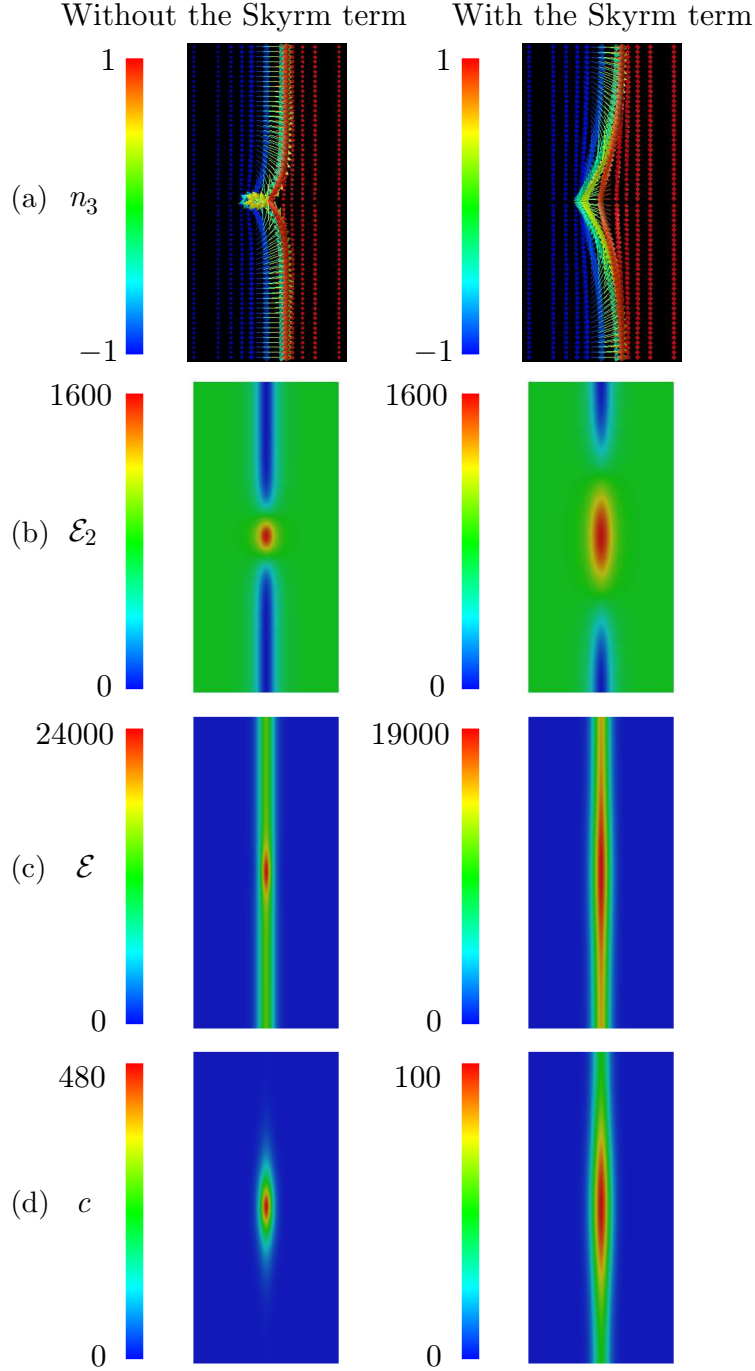}
\caption{
(a) The textures ${\bf n}(x)$.
The color each arrow shows the value of $n_3$.
(b) The energy densities $\mathcal{E}_2 \equiv \beta^2 n_1$.
(c) The total energy densities $\mathcal{E}$.
(d) The topological charge densities $c$.
The left and right columns correspond to the cases without and with the Skyrme term, respectively. 
The numerical box satisfies the periodic boundary condition in the vertical ($x^2$) direction, {\it i.e.}, $n_i(x^1, x^2 + L) = n_i(x^1,x^2)$.
As numerical parameters, we fix $L = 0.5$, $m^2 = 8000$ and $\beta^2 = 800$ for both left and right figures, and $\kappa = 0.002$ for right figures, and plot the values in the region $-0.12 < x^1 < -0.12$ and $0 < x^2 < 0.5$.
\label{fig:wall}}
\end{figure}

In the left column of Fig.~\ref{fig:wall}, 
we give a numerical solution of one sine-Gordon kink 
on the domain wall by using the relaxation method. 
%%%%%%%%%%%%%%%%%%%%%%
%%%%%%%%%%%%%%%%%%%%%%
In (a), we plot our solutions $n_i(x)$ by arrows. 
In (b), we plot the energy contribution from 
the term $\beta^2 n_1$ in the potential, 
in order to show sine-Gordon kinks. 
The total energy density is plot in (c).  
The lump charge density given in the integrand of Eq.~(\ref{eq:lump-charge}) 
is distributed around the sine-Gordon kink 
as seen in (d).
In the right column of Fig.~\ref{fig:wall}, 
we give a numerical solution of the same configuration 
in the presence of the Skyrme term. 
One can see that the size of sine-Gordon kink becomes 
wider due to the Skyrme term.

Multiple sine-Gordon kinks 
repel each other, and such a configuration cannot be static 
on the straight domain wall. 
However, they can be stabilized once the domain wall is closed as 
demonstrated in the next section. 

%%%%%%%%%%%%%%%%%%%%%%%%%%%%%%%%%%%%%%%%%%%%%%%%%%
\section{Jewels on a domain wall ring
\label{sec:wall-ring}}
\begin{figure}%[h]
\centering
\includegraphics[width=0.95\linewidth]{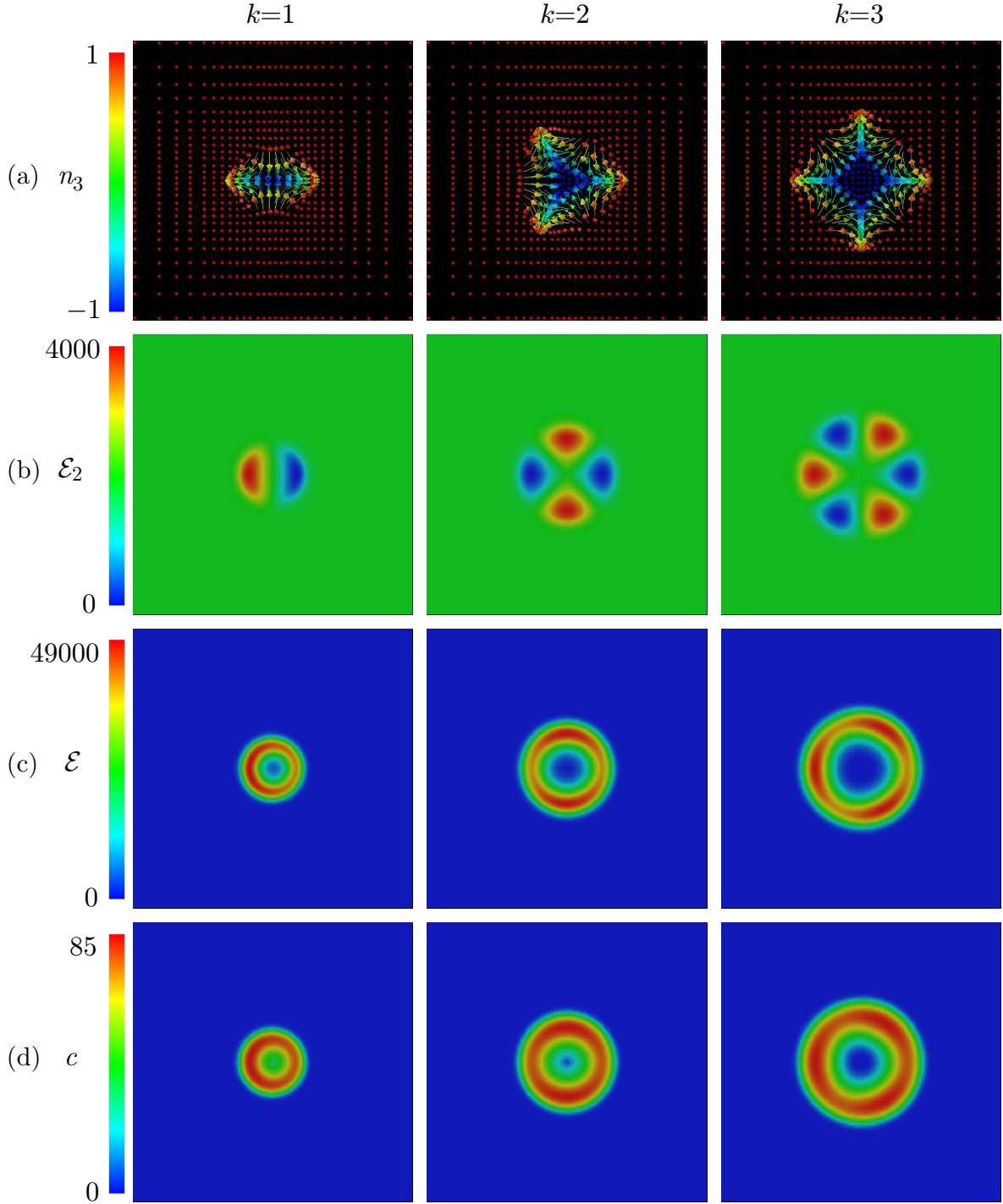}
\caption{
(a) The textures ${\bf n}(x)$.
The color of each arrow shows the value of $n_3$.
(b) The energies $\mathcal{E}_2$.
(c) The total energies $\mathcal{E}$.
(d) The topological charge densities $c$.
The topological charges are $k = 1, 2, 3$ from left to right.
As numerical parameters, we fix $\kappa = 0.02$, $m^2 = 20000$ and $\beta^2 = 2000$,
and plot the values in the region $-0.29 \leq x^{1,2} \leq 0.29$.
\label{fig:wall-ring}}
\end{figure}

Next
we make a closed loop of a domain wall 
with sine-Gordon kinks on it.  
In Fig.~\ref{fig:wall-ring},
we show our numerical results 
by using a relaxation method. 
We constructed configurations with 
the topological lump charge $k=1,2,3$.
In (a), we plot our solutions $n_i(x)$ by arrows. 
In (b), we plot the energy contribution from 
the term $\beta^2 n_1$ in the potential, 
in order to show sine-Gordon kinks.  
One clearly finds that sine-Gordon kinks are separated 
from each other with the same distance for $k=2,3$. 
This is because they repel each other.
In (c), we plot the total energy of each configuration. 
In (d), we plot the topological lump charge density [the integrand of Eq.~(\ref{eq:lump-charge})]. 
One can see that the topological charge density is distributed on the wall ring and 
has peaks at the sine-Gordon kinks. 
%%%%%%%%%%%%%%%%%%%%%%
%%%%%%%%%%%%%%%%%%%%%%

%%%%%%%%%%%%%%%%%%%%%
\section{Summary and Discussion \label{sec:summary} }

We have constructed stable configurations of 
sine-Gordon kinks on a domain wall ring 
in a baby Skyrme model 
with the two potential terms linear and quadratic in fields.
The number of the sine-Gordon kinks coincides with 
the topological lump charge.

Similar configurations of a wall ring with vortices on it are present 
in condensed matter systems, such as 
multi-gap superconductors \cite{Garaud:2012pn} and 
chiral $p$-wave superconductors 
\cite{Garaud:2012} 
in which a four derivative term is not needed.
Our present work was motivated by these works.

Our model can be promoted to a $U(1)$ gauge theory coupled with 
two complex scalar fields $\phi_1(x)$ and $\phi_2(x)$,  
in which lumps are replaced with semi-local vortices.  
In this case, the term $\beta^2 n_1$ is reproduced from 
the Josephson term $\beta^2\phi_1^*\phi_2 +{\rm c.c.}$ 
Then, the model becomes close to exotic superconductors 
considered in \cite{Garaud:2012pn,Garaud:2012}, 
if we replace the Josephson term by
$\beta^2(\phi_1^*\phi_2)^2 +{\rm c.c.}$
In this case, one vortex is decomposed into two fractional vortices 
once absorbed into a domain wall.
However, we still need four derivative term in scalar fields 
for the stability of wall rings. 

If we promote our configuration linearly in $d=3+1$ dimensions,
it becomes a tube with domain lines along it. 
It can be regarded as some exotic cosmic strings 
which may have
some impacts on cosmology.  
For instance, 
it is a very nontrivial question 
whether two of such strings reconnect each other when they collide, because they have internal structures. 
It may be one of the interesting future directions.

\begin{figure}
\includegraphics[width=0.5\linewidth]{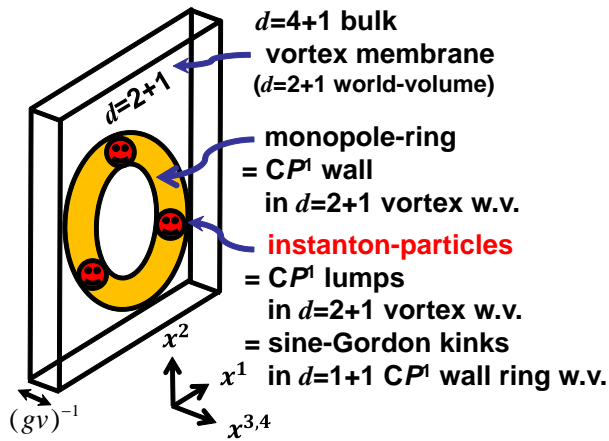}
\caption{Instanton beads on a closed monopole string. \label{fig:instanton}}
\end{figure}

In Ref.~\cite{Nitta:2013cn}, 
a configuration of a sine-Gordon kink on a domain wall 
was embedded into the $2+1$ dimensional world-volume of a non-Abelian vortex \cite{Hanany:2003hp} in $d=4+1$ dimensions. 
In this case, 
the sine-Gordon kinks 
correspond to lumps in the vortex world-volume \cite{Nitta:2013cn}
and to instantons in the bulk \cite{Eto:2004rz,Eto:2006pg},  
while the domain wall in the vortex world-volume 
corresponds to a monopole string in the bulk \cite{Tong:2003pz,Eto:2004rz}. 
The configuration is an instanton confined by 
two monopole strings attached from both sides \cite{Nitta:2013cn}. 
Similarly to this, if one embeds our solution in this paper, 
it becomes instanton beads on a closed monopole string, 
as illustrated in Fig.~\ref{fig:instanton}.
%%%%%%%%%%%%%%%
%%%%%%%%%%%%%%%
For this configuration to be stabilized, one needs higher derivative corrections to the vortex effective action \cite{Eto:2012qda,Nitta:2012mg}.

We have studied a massive ${\bf C}P^1$ model.
One can extend it to massive ${\bf C}P^{n}$ model 
which admits $n-1$ parallel domain walls 
\cite{Gauntlett:2000ib}.
It is an open question whether one can construct 
multiple $n-1$ rings with lump beads on them in this model.
Also, one can discuss non-Abelian domain walls \cite{Isozumi:2004jc} 
in the massive Grassmannian sigma model \cite{Arai:2003tc} 
or in non-Abelian gauge theories.

\section*{Acknowledgements}

M.~N. thanks J.~Garaud for explaining their works 
\cite{Garaud:2012,Garaud:2012pn} 
on vortices within a domain wall ring 
in superconductors, which motivated our work. 
We thank the organizers of the conference 
``Quantized Flux in Tightly Knotted and Linked Systems," 
held in 3 - 7 December 2012 at Isaac Newton Institute 
for Mathematical Sciences, where this work was initiated. 
We thank Paul Sutcliffe for a discussion on 
\cite{footnote}.
This work is supported in part by 
Grant-in-Aid for Scientific Research (Grant No. 22740219 (M.K.) and No. 23740198 (M.N.)) 
and the work of M. N. is also supported in part by 
the ``Topological Quantum Phenomena'' 
Grant-in-Aid for Scientific Research 
on Innovative Areas (No. 23103515)  
from the Ministry of Education, Culture, Sports, Science and Technology 
(MEXT) of Japan.

%%%%%%%%%%%%%%%%%%%%%%%%%%%%%%%%%%%%%%%%%%%%%%%%%%%%%%%%%%%%
%\newpage

%%%%%%%%%% References %%%%%%%%%%%%%%%%%%%%%%%%%
\newcommand{\J}[4]{{\sl #1} {\bf #2} (#3) #4}
\newcommand{\andJ}[3]{{\bf #1} (#2) #3}
\newcommand{\AP}{Ann.\ Phys.\ (N.Y.)}
\newcommand{\MPL}{Mod.\ Phys.\ Lett.}
\newcommand{\NP}{Nucl.\ Phys.}
\newcommand{\PL}{Phys.\ Lett.}
\newcommand{\PR}{ Phys.\ Rev.}
\newcommand{\PRL}{Phys.\ Rev.\ Lett.}
\newcommand{\PTP}{Prog.\ Theor.\ Phys.}
\newcommand{\hep}[1]{{\tt hep-th/{#1}}}
%%%%%%%%%%%%%%%%%%%%%%%%%%%%%%%%%%%%%%%%%%%%%%%

\end{document}